# The Gott-Kaiser-Stebbins (GKS) Effect in an Accelerated Expanding Universe

**S. Y. Rokni**, **H. Razmi**, and **M. R. Bordbar**

Department of Physics, University of Qom, 371614661, Qom, I. R. Iran.

Email(s): razmi@qom.ac.ir & razmiha@hotmail.com; s.y.rokni@gmail.com; mbordbar@qom.ac.ir

**Abstract**

We want to find the cosmological constant influence on cosmic microwave background (CMB) temperature due to moving linear cosmic strings. Considering the space-time metric of a linear cosmic string in an accelerated expanding universe, the Gott-Kaiser-Stebbins (GKS) effect, as an important mechanism in producing temperature discontinuity in the CMB, is considered and its modification due to the effect of the cosmological constant is calculated. The result shows that a positive cosmological constant (i.e. the presence of cosmic strings in an accelerated expanding universe) weakens the discontinuity in temperature so that more strong resolution is needed to detect the corresponding influences on the CMB power spectrum and anisotropy.

**Introduction**

Topological defects have formed during symmetry breaking phase transitions in the early universe [1]. Among different already known defects, scientists have paid most attention to cosmic strings [1-2]. These objects, in addition to their importance in cosmology, have been recently under consideration because of their similarities to fundamental strings [3]. Although one of the first motivations of studying cosmic strings corresponded to density fluctuations due to loop cosmic strings in the early universe as the necessary seeds for the large scale structure we see today [4], the data from COBE and BOOMERANG have shown that cosmic strings cannot be considered as the main candidate for the early density fluctuations in our universe [5]. Indeed, precise measurements in cosmic microwave power spectrum have shown that the most possible contribution of cosmic strings in the early fluctuations can be at most up to 10%; this value constrains the mass density of strings ($G\mu$) up to the limit of $\sim 10^{-7}$ [6]. The recent results from Planck also confirm this constraint ($G\mu < 10^{-7}$) [7].

There are some observational effects for detecting cosmic strings. Loop cosmic strings may be observed via their gravitational radiation [8]; they may be considered as the origin of high-energy cosmic sources [9]. The long (linear) cosmic strings can have gravitational lensing effects [10-11]; they may have some effects on CMB [12-14]. There are possible mechanisms for the contribution of linear cosmic strings in producing temperature discontinuity in CMB [15]; among these, the Gott Kaiser Stebbins (GKS) effect is one of the most important ones [12]. According to this effect, the light ray reaching an observer in front of a linear cosmic string is blue shifted while the ray behind the string remains unchanged; so, the observer sees a small temperature discontinuity due to an angular separation at the order of the string deficit angle. Although for a single cosmic string, it is very hard to observe such an effect, for a network of strings, at small angular scales (high multiples), the GKS effect can have a dominant contribution in angular power spectrum [13].

Considering today observational evidence for an accelerated expanding universe [16], it is natural to study possible modifications of different already known gravitational and cosmological phenomena under the influence of the positive cosmological term $\Lambda$ which is usually considered as the driving force for this acceleration. Here, we want to find the cosmological constant influence on the CMB temperature discontinuity due to cosmic strings (the modified GKS effect).

**A short review of the GKS effect** [17]

Considering the line element of a linear long cosmic string of mass density $\mu$ ($G\mu << 1$) [10]

$$ds^2 = dt^2 - dz^2 - d\rho^2 - (1 - 8G\mu)\rho^2 d\varphi^2 \qquad (1)$$

with the deficit angle $\Delta = 2\pi - \dfrac{\int_0^{2\pi} g_{\phi\phi}^{\frac{1}{2}} d\phi}{\int_0^{\rho} g_{\rho\rho}^{\frac{1}{2}} d\rho} = 8\pi G\mu$, for two particles moving with the same velocity $\vec{v}$ relative to the string, the observed temperature $T_{obs}$ in terms of the background temperature $T_0$ is

$$T_{obs}(\theta) = T_0 \dfrac{(1-u^2)^{\frac{1}{2}}}{(1+u\cos\theta)} \qquad (2),$$

where $u = 2v\sin\dfrac{\Delta}{2}$ and $\cos\theta = +1$ ($\cos\theta = -1$) for when the source of emission of the observed photons and the observer are moving away from (toward) each other. The corresponding temperature discontinuity is simply found as:

$$\frac{\delta T(\theta)}{T_0} = \frac{T_{obs} - T_0}{T_0} \approx -u\cos\theta + \frac{u^2}{2}\left(\cos(2\theta)\right) + O(u^3) \qquad (3);$$

Or:

$$\frac{\delta T}{T_0} = 8\pi\gamma(v)vG\mu \qquad (4),$$

where $\gamma(v)$ is the Lorentz relativistic gamma factor which is about 1 ($\gamma(v) \approx 1$) for when $v \ll c$. The relation (4) is known as the GKS effect.

**Cosmological constant influence on the GKS effect**

Using the following already known line element around a linear long cosmic string under the influence of a positive cosmological constant $\Lambda$ [18]

$$ds^2 = \cos^{\tfrac{4}{3}}\!\left(\frac{\sqrt{3\Lambda}}{2}\rho\right)(dt^2 - dz^2) - d\rho^2 - \frac{4(1-4G\mu)^2}{3\Lambda}\cos^{\tfrac{4}{3}}\!\left(\frac{\sqrt{3\Lambda}}{2}\rho\right)\tan^2\!\left(\frac{\sqrt{3\Lambda}}{2}\rho\right)d\varphi^2 \qquad (5),$$

the deficit angle, the relative velocity, and the temperature discontinuity are modified as:

$$\Delta = 2\pi - 2\pi \left[ \frac{2(1-4G\mu)}{\sqrt{3\Lambda}\rho} \cos^{-\frac{1}{3}}\left(\frac{\sqrt{3\Lambda}}{2}\rho\right) \sin\left(\frac{\sqrt{3\Lambda}}{2}\rho\right) \right]$$
$$\approx 8\pi G\mu - \frac{1}{40}\pi\Lambda^2\rho^4 + \frac{1}{40}\pi\Lambda^2\rho^4 G\mu \tag{6}$$

$$u = \gamma(v)(v\sin\frac{\Delta_{sl}}{2} + v\sin\frac{\Delta_{so}}{2}) \approx \gamma(v)v(8\pi G\mu - \frac{1}{80}\pi\Lambda^2\rho_{sl}^{4} - \frac{1}{80}\pi\Lambda^2\rho_{so}^{4}) \tag{7}$$

and

$$\frac{\delta T}{T_0} = 8\pi\gamma(v)vG\mu(1 - \frac{1}{640}\frac{1}{G\mu}\Lambda^2(\rho_{sl}^{4} + \rho_{so}^{4})) \tag{8}$$

where $\rho_{sl}$ and $\rho_{so}$ are the distances from the string to the last scattering surface and the observer respectively.

The relation (8) can be considered as the modified GKS effect.

**Conclusion**

Although $\Lambda$ has a very small value ($\sim 10^{-52} m^{-2}$), for cosmic scales values of $\rho_{sl}$ and $\rho_{so}$ ($\leq R_H \sim 10^{25} m$) and because of small value of $G\mu(<10^{-7})$, the modification term $\frac{1}{640}\frac{1}{G\mu}\Lambda^2(\rho_{sl}^{4} + \rho_{so}^{4})$ in (8) not only isn't negligible but also may be comparable to *1*. This means the modification of the GKS effect can weaken the standard GKS effect considerably. Among other things, an important consideration is that one needs more strong resolution to detect the discontinuity in the CMB temperature due to cosmic strings. Therefore, considering current observational apparatuses, it may take a long time to be able to detect cosmic strings. As we know, a cosmic string alone cannot affect on the CMB; but, it is a set of these topological defects (e.g. a network) which can dominantly affect the CMB anisotropy power spectrum for larger values of the orders of the spherical harmonics multiple expansion ($l > 3000$) [19]. It seems it is necessary to reconsider the CMB anisotropy power spectrum based on the relation (8).

Finally, it is good to checking the special limiting case $G\mu \ll \Lambda^2 \rho_{sl,s0}^4$ which may occur for when $G\mu \ll 10^{-7}$:

$$\frac{\delta T}{T_0} \approx -\frac{\pi \gamma(v)v}{80}\Lambda^2(\rho_{sl}^4 + \rho_{so}^4) \qquad (9).$$

The negative definite value of this result is justified based on this fact that $\Lambda$ acts as an "antigravity" force.